\documentclass[aps,pre,showpacs,amsmath,amsfonts,amssymb,superscriptaddress,twocolumn,floatfix]{revtex4-1} 
\usepackage[dvips]{graphicx}
\usepackage{bbm}
\usepackage[applemac]{inputenc}
\newcommand{\beq}{\begin{equation}} 
\newcommand{\eeq}{\end{equation}}
\newcommand{\bea}{\begin{eqnarray}} 
\newcommand{\eea}{\end{eqnarray}}

\usepackage{xcolor}
\usepackage[colorlinks,linkcolor=blue,anchorcolor=blue,citecolor=blue,filecolor=blue,menucolor=blue,urlcolor=blue,breaklinks]{hyperref}
\usepackage[hyphenbreaks]{breakurl}

\input epsf 
 
\begin{document} 
 
\title{Temperature inversion in long-range interacting systems} 

\author{Tarc\'{\i}sio N.\ Teles} 
\email{tarcisio.nteles@gmail.com} 
\thanks{S.\ Gupta and T.\ N.\ Teles contributed equally to the work.}
\affiliation{\mbox{Instituto de F\'{\i}sica, Universidade Federal do Rio
Grande do Sul, Caixa Postal 15051, CEP 91501-970, Porto Alegre, RS,
Brazil}}  
\affiliation{Departamento de Ciências Exatas e Sociais Aplicadas, Universidade Federal de Ciências da Saúde de Porto Alegre, CEP 90050-170, Porto Alegre, RS, Brazil}

\author{Shamik Gupta}
\email{shamikg1@gmail.com} 
\affiliation{\mbox{Dipartimento di Fisica e Astronomia and CSDC, Universit\`a di Firenze, and INFN, sezione di Firenze,} via G.\ Sansone 1, I-50019 Sesto Fiorentino, Italy}  

\author{Pierfrancesco Di Cintio} 
\email{pierfrancesco.dicintio@unifi.it} 
\affiliation{\mbox{Dipartimento di Fisica e Astronomia and CSDC, Universit\`a di Firenze, and INFN, sezione di Firenze,} via G.\ Sansone 1, I-50019 Sesto Fiorentino, Italy}  

\author{Lapo Casetti} 
\email{lapo.casetti@unifi.it} 
\affiliation{\mbox{Dipartimento di Fisica e Astronomia and CSDC, Universit\`a di Firenze, and INFN, sezione di Firenze,} via G.\ Sansone 1, I-50019 Sesto Fiorentino, Italy}  
\affiliation{INAF-Osservatorio Astrofisico di Arcetri, Largo E.\ Fermi 5, I-50125 Firenze, Italy}  

\date{\today} 
 
\begin{abstract} 
Temperature inversions occur in nature, e.g., in the solar corona and in interstellar molecular clouds: somewhat counterintuitively, denser parts of the system are colder than dilute ones. 
We propose a simple and appealing way to \textit{spontaneously} generate temperature inversions in systems with long-range interactions, by preparing them in inhomogeneous thermal equilibrium states and then applying an impulsive perturbation. In similar situations, short-range systems would typically relax to another thermal equilibrium, with uniform temperature profile. By contrast, in long-range systems, the interplay between wave-particle interaction and spatial inhomogeneity drives the system to nonequilibrium stationary states that generically exhibit temperature inversion. We demonstrate this mechanism in a simple mean-field model and in a two-dimensional self-gravitating system. Our work underlines the crucial r\^{o}le the range of interparticle interaction plays in determining the nature of steady states out of thermal equilibrium.
\end{abstract} 
 
 
 
\maketitle 

Stationary states far from thermal equilibrium occur in nature. In some cases, e.g., in the solar corona and in interstellar molecular clouds, such states exhibit \textit{temperature inversion}: denser parts of the system are colder than dilute ones. This work is motivated by an attempt to explain how such a counterintuitive effect may spontaneously arise in nonequilibrium states, unveiling its minimal ingredients and the underlying physical mechanism. To this end, we start with asking a simple yet physically relevant question: What happens if an isolated macroscopic system in thermal equilibrium is momentarily disturbed, e.g., by an impulsive force or a ``kick''? If the interactions among the system constituents are short-ranged, collisions redistribute the kick-injected energy among the particles, yielding a fast relaxation to a new equilibrium, with a Maxwellian velocity distribution and a uniform temperature across the system.

Is the scenario the same if instead the interactions are long-ranged \footnote{Interactions are called long-range when they decay asymptotically with the interparticle distance $r$ as $r^{-\alpha}$ with $0 \leq \alpha \leq d$ in $d$ dimensions \cite{CampaEtAl:book}.}? For long-range systems, collisional effects act over a characteristic time $\tau_{\text{coll}}$ that, unlike short-range systems, scales with the system size $N$, diverging as $N \to \infty$ \cite{CampaEtAl:book}. As a result, a macroscopic system with long-range interactions starting from generic initial conditions will attain thermal equilibrium only after extremely long times, often exceeding typical observation times. Examples of long-range systems are self-gravitating systems, for which, e.g., $\tau_{\text{coll}} \simeq 10^{10}$ years for globular clusters and orders of magnitude larger than the age of the universe for galaxies \cite{Choudhouri:book,BinneyTremaine:book}. The collisionless evolution of long-range interacting systems for times shorter than $\tau_{\text{coll}}$ is governed by the Vlasov (or collisionless Boltzmann) equation \cite{CampaEtAl:book}. When kicked out of thermal equilibrium, a long-range interacting system relaxes to a Vlasov-stationary state, and thermal equilibrium is just one of infinitely many possible states. Predicting which Vlasov state follows a given initial condition is an extremely difficult problem \footnote{Also the related issue of \textit{stability} of the Vlasov stationary states is resolved only in particular cases \cite{CampaEtAl:book}.}, first addressed in \cite{Lynden-Bell:mnras1967} and still unsolved, despite some remarkable achievements for particular systems and special initial conditions (see \cite{LevinEtAlphysrep:2014,BenettiEtAl:prl2014}). 

Let us then ask a simpler question: How different from an equilibrium state
is the stationary state the system relaxes to after the kick? Are there ways to
characterize it, e.g., by unveiling some of its general features? The answer is
yes; in this paper, we argue that, provided it started from a spatially
inhomogeneous equilibrium state \footnote{At variance with short-range
systems, equilibrium states of long-range systems can be
inhomogeneous, typically when the interactions are attractive.}, the
system after the kick relaxes to a state with a non-uniform temperature
profile.
In short-range systems, by contrast, a
non-uniform temperature profile may only occur when the system is
actively maintained out of equilibrium, e.g., by a boundary-imposed
temperature gradient, to counteract collisional effects. Remarkably, in
a long-range system, the relaxed state after the kick generically exhibits temperature inversion, as we will explicitly show. 

As recalled above, temperature inversions are observed \footnote{Temperature inversions
have been observed also in other systems like the hot gas of
``cooling-core'' galaxy clusters \cite{WiseMcNamaraMurray:apj2004} and the Io plasma torus around Jupiter \cite{MeyerVernetMoncuquetHoang:icarus1995}.} in interstellar molecular clouds \cite{MyersFuller:apj1992,GoodmanEtAl:apj1998,PinedaEtAl:apjl2010} and especially in the solar corona, where temperatures around $10^6$ K that are
three orders of magnitude larger than the temperature of the photosphere
are attained \cite{GolubPasachoff:book}. Despite recent advances
\cite{Aschwanden:book}, the mechanism of coronal heating is not
completely understood and remains one of the most important open
problems in astrophysics \cite{Klimchuk:sp2006}. Most attempts to
explain such a phenomenon involve mechanisms that actively inject
energy \footnote{By means of, e.g., magnetic reconnection or Alfv\'en
waves \cite{Klimchuk:sp2006}.} into the less dense regions of
the system. A different possibility, suggested by Scudder
\cite{Scudder:apj1992a,Scudder:apj1992b,Scudder:apj1994}, is
referred to as velocity filtration; see also Ref.\ \cite{epjb2014} and Appendix \ref{appendix_supplmat}. Consider a system of particles acted
upon by an external field whose potential energy increases with height
above a base level. Only particles with a sufficiently large kinetic
energy can climb the potential well and reach a given height. If the
velocity distribution at the base is Maxwellian, it remains as such with 
the system maintaining the same temperature at all heights in the
stationary state. Instead, if the distribution is suprathermal, i.e., with tails fatter than Maxwellian, the temperature in the stationary state increases with height, with concomitant decrease of the density. It is argued \cite{Klimchuk:sp2006,Anderson:apj1994,LandiPantellini:aa2001} that velocity filtration might not be the (only) mechanism behind coronal heating, but nevertheless provides a simple and appealing explanation of how a counterintuitive temperature inversion occurs without steady energy injection in less dense parts of the system.  

Scudder's original model neglects interparticle interactions and requires 
an ``active'' ingredient, i.e., an out-of-equilibrium suprathermal
velocity distribution imposed as a boundary condition. It was recently
shown \cite{epjb2014} that temperature inversion occurs also in strongly
interacting systems, provided the interactions are long-ranged, when a
velocity distribution with suprathermal tails is given just as the
initial condition of the dynamical evolution. Although much weaker than
a nonthermal boundary condition, the latter is still an \textit{ad hoc}
requirement. However, it is not necessary at all: in this paper, we demonstrate that temperature inversion emerges \textit{spontaneously} in the stationary state reached after a long-range system is brought out of equilibrium by a perturbation acting for a very short time, and there is no need for a suprathermal initial distribution. Our work thus suggests that temperature inversions observed in astrophysical systems may be examples of a more general phenomenon, whose roots are in the long-range nature of the interparticle interactions.

Let us consider the very general setting of a system of $N$ interacting particles of mass $m$ in $d$ dimensions with Hamiltonian
\beq
\mathcal{H} = \sum_{i=1}^{N} \frac{p_i^2}{2m} + \sum_{i = 1}^N \sum_{j > i}^N V \left(\left|\mathbf{r}_i - \mathbf{r}_j \right|\right)\,,
\label{genericH}
\eeq
where $p_i = \left|\mathbf{p}_i \right|$ are the momenta conjugated to
the positions $\mathbf{r}_i$, the potential energy is long-ranged, $V(r)
\propto r^{-\alpha}; 0\leq \alpha \leq d ~\text{as}~ r\to \infty$. For times $t \ll \tau_{\text{coll}}$, the dynamics is described by the single-particle phase space distribution $f(\mathbf{r},\mathbf{p},t)$ obeying the Vlasov equation
\beq
\frac{\partial f}{\partial t} + \mathbf{p} \cdot \nabla_{{\mathbf{r}}}f
- \nabla_{{\mathbf{r}}} u\left[f\right]  \cdot \nabla_{{\mathbf{p}}}f = 0\, , 
\label{vlasov}
\eeq 
where
\beq
u\left[f\right] \equiv \int d\mathbf{r}\, d\mathbf{p}\, f(\mathbf{r},\mathbf{p},t)\, V(r)
\label{mfv}
\eeq  
is the mean-field potential energy. An initial condition $f_0=
f(\mathbf{r},\mathbf{p},0)$ chosen to be a stationary solution of Eq.\
\eqref{vlasov} does not evolve in time. Otherwise, if the initial distribution is not a stationary solution of \eqref{vlasov}, after a short
transient (often referred to as ``violent relaxation'', after
Lynden-Bell \cite{Lynden-Bell:mnras1967}), the system settles into a
stable stationary solution of Eq.\ \eqref{vlasov} called a
quasi-stationary state (QSS), in which the system remains trapped until, at $t \simeq
\tau_{\text{coll}}$, collisional effects neglected in Eq.\
\eqref{vlasov} drive the system towards thermal equilibrium
\cite{CampaEtAl:book}. To construct a first representative example to demonstrate our claims, we assume periodic coordinates so that boundary
effects may be neglected \footnote{At variance with Scudder's model we do not need any special boundary condition.} and expand the interparticle potential (that by
definition is even) in a cosine Fourier series. We then set $d=1$ and retain just the first Fourier term. The resulting model, the so-called Hamiltonian Mean-Field (HMF) model, describes a system of globally interacting point particles moving on a circle with Hamiltonian \cite{AntoniRuffo:pre1995}
\beq
\mathcal{H}_{\text{HMF}} = \sum_{i=1}^{N} \frac{p_i^2}{2} + \frac{1}{N}\sum_{i = 1}^N \sum_{j > i}^N  \left[ 1 - \cos \left(\vartheta_i - \vartheta_j \right) \right]\,,
\label{hmf}
\eeq
where $\vartheta_i\in (-\pi,\pi]$ is the angular coordinate of the $i$th
particle on the circle, while $p_i$ is the conjugated
momentum; we have further assumed $m = 1$ and the interaction to be attractive, fixing the energy scale to unity, and scaled it by $1/N$ to make it extensive (Kac prescription,
\cite{CampaEtAl:book}).
The Hamiltonian \eqref{hmf} is invariant under the O$(2)$ symmetry
group. In thermal equilibrium, for energy density $\varepsilon =
E/N$ smaller than $\varepsilon_c = 3/4$, the symmetry is
spontaneously broken to result in a clustered state. The order
parameter of clustering is the average magnetization \footnote{The HMF
may also be seen as a system of mean-field $XY$ spins.} $(m_x,m_y) \equiv \frac{1}{N}\left( \sum_{i=1}^N \cos \vartheta_i, \sum_{i=1}^N \sin
\vartheta_i \right)$.
The HMF model is a simple system which, besides serving as a framework to study statics and dynamics of long-range systems, actually models physical systems like gravitational sheet models and free-electron lasers \cite{CampaEtAl:book}. 

\begin{figure}
\centerline{\includegraphics[width=55mm]{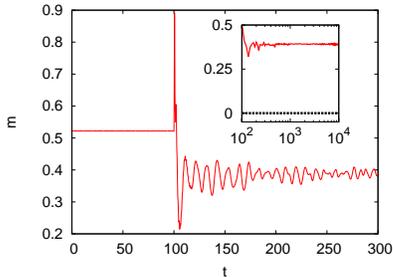}}
\caption{(Color online) HMF: time evolution of the magnetization $m$ with $N
=10^7$ (solid red line). Inset: $m(t)$ compared to the equilibrium value $m_{\text{eq}} = 0$ (dotted black line) for longer times.}
\label{fig:m(t)}
\end{figure}

\begin{figure}
\centerline{\includegraphics[width=55mm]{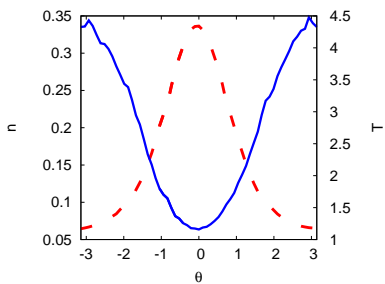}}
\caption{(Color online) HMF: temperature profile $T(\vartheta)$ (blue
solid line) and density profile $n(\vartheta)$ (red dashed line), measured in the QSS at $t = 10^4$.}
\label{fig:temp}
\end{figure}

\begin{figure}
\centerline{\includegraphics[width=55mm]{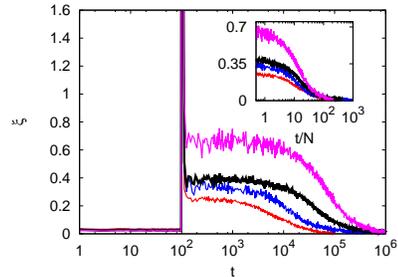}}
\caption{(Color online) HMF: Time evolution of the distance from
equilibrium temperature $\xi$, Eq.\ \eqref{eq:xi}, with $N$ increasing from bottom to top: $N = 5\times10^2$ (red), $N = 10^3$ (blue), $N = 2.5\times10^3$ (black),  $N = 5\times 10^3$ (purple). Each curve is the average over $n_r$ realizations with $n_r$ ranging from $20$ for $N = 5\times 10^3$ to $10^3$ for $N = 5\times 10^2$. Inset: $\xi$ as a function of $t/N$.}
\label{fig:xi}
\end{figure}

In order to study what happens when we ``kick'' the HMF system out of equilibrium, we studied its dynamical evolution via molecular dynamics (MD) simulations, performed by integrating \footnote{We used a $4^{\text{th}}$ order symplectic algorithm with time step $\delta t =0.1$, keeping relative energy fluctuations below $10^{-8}$.} the equations of motion derived from the Hamiltonian (\ref{hmf}).
We prepared the system in an equilibrium state
with $m_x = m_0$ and $m_y = 0$, and with a Maxwellian velocity
distribution corresponding to the equilibrium temperature. We let the
system evolve until $t = t_0>0$, and then kicked it out of equilibrium
by applying during a short time $\tau$ an external magnetic field $h$ along the $x$ direction;
thus, for $t_0 < t < t_0 + \tau$, the Hamiltonian \eqref{hmf} is augmented by
$\mathcal{H}_h = - h\sum_{i=1}^N \cos \vartheta_i$.
Here we present results for $t_0 = 100$, $\tau = 1$, $h = 10$, and $m_0 = 0.521$, corresponding to an initial equilibrium temperature $T = 0.4244$. We considered up to $N = 10^7$ particles. 
After the kick the magnetization starts oscillating, and after a transient the oscillations damp down and a new stationary value $m^* < m_0$ of the magnetization is reached. A typical time evolution of the magnetization is shown in Fig.\ \ref{fig:m(t)}. 
The fact that $m^* < m_0$ is not surprising because the system gains
energy during the interaction with the external field $h$, resulting in an energy density $\varepsilon^* > \varepsilon_c$, and if the system would
relax to a new equilibrium, the latter would have a magnetization $m_{\text{eq}} = 0$. The stationary state reached after the damping of the oscillations is a QSS, very far from the homogeneous equilibrium at $\varepsilon^*$. The nonequilibrium character of this state is further shown by the fact that the temperature profile
\beq
T(\vartheta) \equiv \frac{\int_{-\infty}^\infty dp\, p^2 f(\vartheta,p)}{\int_{-\infty}^{\infty} dp\, f(\vartheta,p)}
\label{tempprofile}
\eeq
is non-uniform, and there is temperature inversion, as shown in Fig.\ \ref{fig:temp}, where $T(\vartheta)$ is plotted together with the density profile 
\beq
n(\vartheta) \equiv \int_{-\infty}^{\infty} dp\, f(\vartheta,p)\,.
\label{densprofile}
\eeq 
The temperature profile indeed
remains essentially the same for the whole lifetime of the QSS, as we
checked by measuring an integrated distance $\xi$ between the
actual temperature profile and the constant equilibrium one,
$T_{\text{eq}}$, at the same energy, as follows:
\beq
\xi(t) \equiv \int_{-\pi}^\pi \left| T(\vartheta,t) - T_{\text{eq}} \right| d\vartheta \,.
\label{eq:xi}
\eeq
In Fig.\ \ref{fig:xi}, $\xi(t)$ is plotted for systems with different
values of $N$ kicked with the
same $h = 10$ at $t_0 = 100$ for a duration $\tau = 1$. After the kick, $\xi(t)$ oscillates and then
reaches a plateau whose duration grows with $N$, as expected for a QSS.
The inset of Fig.\ \ref{fig:xi} shows that if times
are scaled by $N$, the curves reach zero at the same time, consistently
with the lifetime of an inhomogeneous QSS being proportional to $N$ \cite{deBuylMukamelRuffo:pre2011}. 

We performed simulations starting with equilibrium states corresponding
to different values of $m_0$ and applying perturbations of different
strengths $h$ and duration $\tau$, and the system almost always ended up in a QSS showing at least a partial temperature inversion over some interval of the values of
$\vartheta$, if not the whole; we also considered another model where also the second Fourier mode is retained, again obtaining the same qualitative behaviour (see Appendix \ref{appendix_supplmat}). 

\begin{figure}
\centerline{\includegraphics[width=55mm]{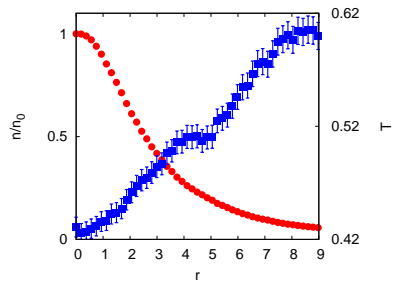}}
\caption{(Color online) 2DSGS: temperature profile $T(r)$ (blue squares) and density  profile $n(r)/n_0$ (red circles) measured in the
QSS at $t = 90\,\tau_{\text{dyn}}$ while starting from a thermal state with uniform temperature $T_0 = 0.5$, in natural units such that $r_*(t) \simeq 7$ and $r_0 = 10^{-3}$. 
}
\label{fig:2dgravity}
\end{figure}

Preparing the system in equilibrium and then bringing it out of
equilibrium by means of a perturbation acting for a short time mimics
processes that may actually happen in nature, e.g., a transient
density perturbation in a self-gravitating fluid. Hence, as a second example we considered a two-dimensional self-gravitating system (2DSGS), that is, a system of $N$ particles of mass $m$ moving in a plane with Hamiltonian 
\beq
\mathcal{H}_{\text{G}} = \sum_{i = 1}^N \frac{\left|\mathbf{p}_i\right|^2}{2m} + Gm^2
\sum_{i, j > i}^N  
\ln\left[\frac{\sqrt{(\mathbf{r}_i - \mathbf{r}_j)^2 + r_0^2}}{s}\right]\,,
\label{ham_2dgrav}
\eeq
where $\mathbf{r}_i = (x_i,y_i)$, $s$ is a length scale, $\mathbf{p}_i = (\dot{x}_i,\dot{y}_i)$, $r_0$ is a small-scale cutoff 
and $G$ is the gravitational constant. Such a system can be seen as a simple model of filamentary interstellar clouds \cite{TociGalli:mnras2015}. We performed MD simulations considering $N = 3 \times 10^4$ particles initially in a thermal equilibrium state, whose radial density profile is known \cite{Ostriker:apj1964} (see Appendix \ref{appendix_supplmat}), and we kicked them out of equilibrium by instantaneously adding to all their radial velocities the same amount $\delta v_r = \sigma_r/2$, where $\sigma_r$ is the radial velocity dispersion. 
As in the HMF case, after the kick the system develops macroscopic oscillations, that damp out after a time of order $\tau_{\text{dyn}} = r^0_*\sqrt{2/GM}$ where $M$ is the total mass and $r^0_*=r_*(t=0)$ is the (initial) half-mass radius, and eventually sets in a QSS. The latter exhibits temperature inversion up to $ r \gtrsim r_*(t)$, as shown in Fig.\ \ref{fig:2dgravity}. The radial profiles $n(r)$ and $T(r)$ are obtained by averaging $T(\mathbf{r})$ and $n(\mathbf{r})$, defined by replacing $(\vartheta,p)$ with $(\mathbf{r},\mathbf{p})$ in Eqs.\ \eqref{tempprofile} and \eqref{densprofile}, over the polar angle. 

We have thus shown that temperature inversion arises in two different long-range-interacting systems when kicked out of equilibrium. Moreover, also if we start from a state with a Maxwellian velocity distribution but a spatial distribution different from the equilibrium one, macroscopic oscillations develop, and after the damping of the oscillations, the systems end up in a QSS typically exhibiting at least a partial temperature inversion (see Appendix \ref{appendix_supplmat}). Therefore, this phenomenon does not depend on the details of the way the initial state is prepared, provided it is inhomogeneous, and that a collective oscillation develops when it is brought out of equilibrium.

A very general mechanism may then be responsible for this
phenomenology. In the following, we argue that indeed a simple and
general explanation can be found. As stated above, before collisional
effects set in, the dynamics of a long-range interacting system is described by the Vlasov
equation \eqref{vlasov}.
When a state described by a stationary solution of the Vlasov equation
is perturbed, a phenomenon called Landau damping
\cite{Choudhouri:plasmabook} occurs, which is a kind of wave-particle
interaction responsible for the collisionless damping of oscillations
induced by the external perturbation. The theory of Landau damping is
well developed, and, in the case of small perturbations of homogeneous
states, has been recently put on rigorous grounds
\cite{MohoutVillani:actamath2011}. The phenomenon occurs also when the
perturbed state is inhomogeneous
\cite{BarreOlivettiYamaguchi:jstat2010,BarreOlivettiYamaguchi:jphysa2011}.
The basic physical mechanism is the following: The perturbation creates
a wave in the system, as witnessed by the onset of collective oscillations. Consider for simplicity a monochromatic wave. Particles that have a velocity $v$ close to the phase velocity $v_{\text{ph}}$ of the wave will interact strongly with the perturbation, while those with  $v \ll v_{\text{ph}}$ or $v \gg v_{\text{ph}}$ will be essentially unaffected. Particles with velocities slightly less than $v_{\text{ph}}$ will acquire kinetic energy, while those with velocities slightly larger than $v_{\text{ph}}$
will lose kinetic energy. Since for a typical $f(v)$ one has $df/dv <
0$ (for positive $v$, and the opposite for negative $v$), there are more
particles that gain energy than those that lose energy, so that the
wave in effect loses energy, and the perturbation is damped. After the interaction with the perturbation, a Maxwellian $f(v)$ is modified close to $v_{\text{ph}}$ (and if the wave is stationary or travels in both directions, as in our case, also close to $-v_{\text{ph}}$), where a ``shoulder'' is formed, and $\langle v^2 \rangle$ is
increased. If, prior to
perturbation, the state is spatially homogeneous, this happens in the
same way throughout the system, and the initially coherent energy of the
wave goes into uniformly heating the system. But in our case the state
is \textit{not} homogeneous prior to the perturbation: it is clustered,
and the ``shoulder'' in $f(v)$ created by Landau damping is nothing but a suprathermal tail, so that now velocity filtration acts and produces temperature inversion:
fast particles climb the potential well higher than slower ones, and
suprathermal tails grow when density decreases. Indeed, as shown in
\cite{epjb2014}, velocity filtration always works in systems described
by the Vlasov equation \eqref{vlasov}; there, the suprathermal tails
of $f(v)$ were fed by the initial conditions, while here it is the
perturbation that creates them, via Landau damping. It is interesting
to note that standing radiofrequency waves are used to heat laboratory
plasmas by creating a shoulder in the velocity distribution  \cite{BowersBirdsall:physplasmas2002}; in that case, one needs to
sustain the wave from outside, while in our
example it is velocity filtration that amplifies the effect.

\begin{figure*}
\centerline{\includegraphics[width=110mm]{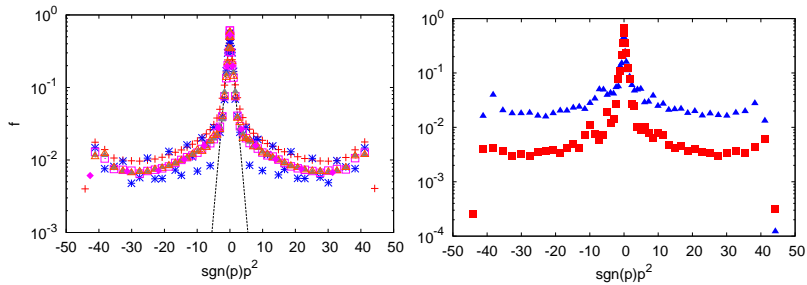}}
\caption{(Color online) HMF: (left) Momentum distribution $f(p)$ at $t=0$
(black dashed line), $t = 101$ (red plus signs), $t = 150$ (magenta empty squares), $t = 200$ (grey filled triangles), $t = 10^3$ (orange empty triangles), and $t = 10^4$ (magenta filled rhombi). (Right) Distribution function $f(\vartheta,p)$ measured in the QSS as in Fig.\ \ref{fig:temp}, at $\vartheta = 0$ (red squares) and $\vartheta = \pi$ (blue triangles). The distribution functions are plotted against $p^2 \text{sgn}(p)$ to better show the difference with respect to the initial Maxwellian.}
\label{fig:veldistrtime}
\end{figure*}

The above physical picture is idealized and one should take into account
the coupling of the various modes of the perturbation in the
inhomogeneous case
\cite{BarreOlivettiYamaguchi:jstat2010,BarreOlivettiYamaguchi:jphysa2011}.
This notwithstanding, the main point is that the perturbation does not
interact in the same way with all the particles, but preferentially
gives energy to particles with already rather large velocities. This
results in a suprathermal velocity distribution, as shown in Fig.\
\ref{fig:veldistrtime} (left panel) for the HMF model. Soon after the kick a high-$p$ tail shows up, which is built up from peaks at different values of $p$ corresponding to different positions. The evolution of these peaks results in oscillations in $f(p)$ for $t \lesssim 150$, then the oscillations damp out and due to the absence of an efficient mechanism able to evenly redistribute this excess energy among all the particles, the velocity distribution stays nonthermal and essentially the same for times $t < \tau_{\text{coll}}$, allowing velocity filtration to produce temperature inversion. The distribution function of the HMF model in the QSS is plotted for two different values of $\vartheta$ in Fig.\ \ref{fig:veldistrtime} (right panel), and the growth of the suprathermal tails in the less dense parts of the system is well apparent.  
We found similar results for the distribution functions also in the 2DSGS case (see Appendix \ref{appendix_supplmat}). For the HMF, perturbations of the form $-h\sum_{i=1}^N \cos\left(k \vartheta_i\right)$ with $k \not = 1$ also yield preferential absorption of energy around values of $v$ that are different from those in the $k = 1$ case, coherently with the above picture (see Appendix \ref{appendix_supplmat}). 

We have shown that nonequilibrium stationary states with temperature inversions are the typical outcome of a perturbation acting for a short time on a clustered equilibrium state of a long-range interacting system. This rather surprising result can be explained in terms of Landau damping and velocity filtration, suggesting that temperature inversions  may occur whenever the dynamics is collisionless up to the relevant timescales. 
This mechanism may be actually relevant to understand temperature inversions observed in nature.

\textit{Acknowledgments.} 
We thank the Galileo Galilei Institute for Theoretical Physics, Florence, Italy for the hospitality and the INFN for partial support during the completion of this work. LC thanks D.\ Galli for very useful discussions.

\appendix

\section{Supplemental material}
\label{appendix_supplmat}

In this Appendix we present some supplemental material and in particular, report on the results of molecular dynamics simulations performed using different protocols and sets of parameters with respect to those shown in the paper. Moreover, we show some further details and results on the two-dimensional self-gravitating system and finally we present a tutorial discussion of the mechanism of velocity filtration.

\subsection{HMF model}

Let us start with results for the HMF model, whose Hamiltonian is  
\beq
\mathcal{H}_{\text{HMF}} = \sum_{i=1}^{N} \frac{p_i^2}{2} + \frac{1}{N}\sum_{i = 1}^N \sum_{j > i}^N  \left[ 1 - \cos \left(\vartheta_i - \vartheta_j \right) \right]\,,
\label{hmfApp}
\eeq
where $\vartheta_i\in (-\pi,\pi]$ is the angular coordinate of the $i$th
particle and $p_i$ is the conjugated
momentum. The system is prepared in a thermal equilibrium state with temperature $T$ and magnetization $m=m_0$, where $m$ is the modulus of the vector 
\beq
(m_x,m_y) \equiv \frac{1}{N}\left( \sum_{i=1}^N \cos \vartheta_i, \sum_{i=1}^N \sin
\vartheta_i \right)\,. 
\label{eq:mxmy}
\eeq
After a time $t_0 > 0$, an interaction with a magnetic field $h$ is switched on by adding to the Hamiltonian a term 
\beq
\mathcal{H}_h \equiv - h\sum_{i=1}^N \cos \vartheta_i 
\label{eq:pert}
\eeq
and then switched off again at a time $t_0 + \tau$.

\subsubsection{Density and temperature profiles}

\begin{figure}
\centerline{\includegraphics[width=65mm]{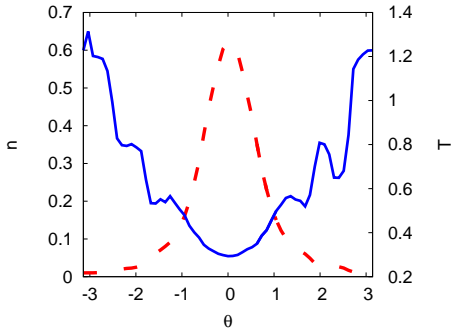}}
\caption{HMF: Temperature profile $T(\vartheta)$ (blue solid
line) and density profile $n(\vartheta)$ (red dashed line) measured in the
QSS at $t = 10^4$, in the same
conditions as Fig.\ 1 of the article, with the difference that $T=0.25$.}
\label{fig-sm:temp0-25}
\end{figure}

\begin{figure}
\centerline{\includegraphics[width=65mm]{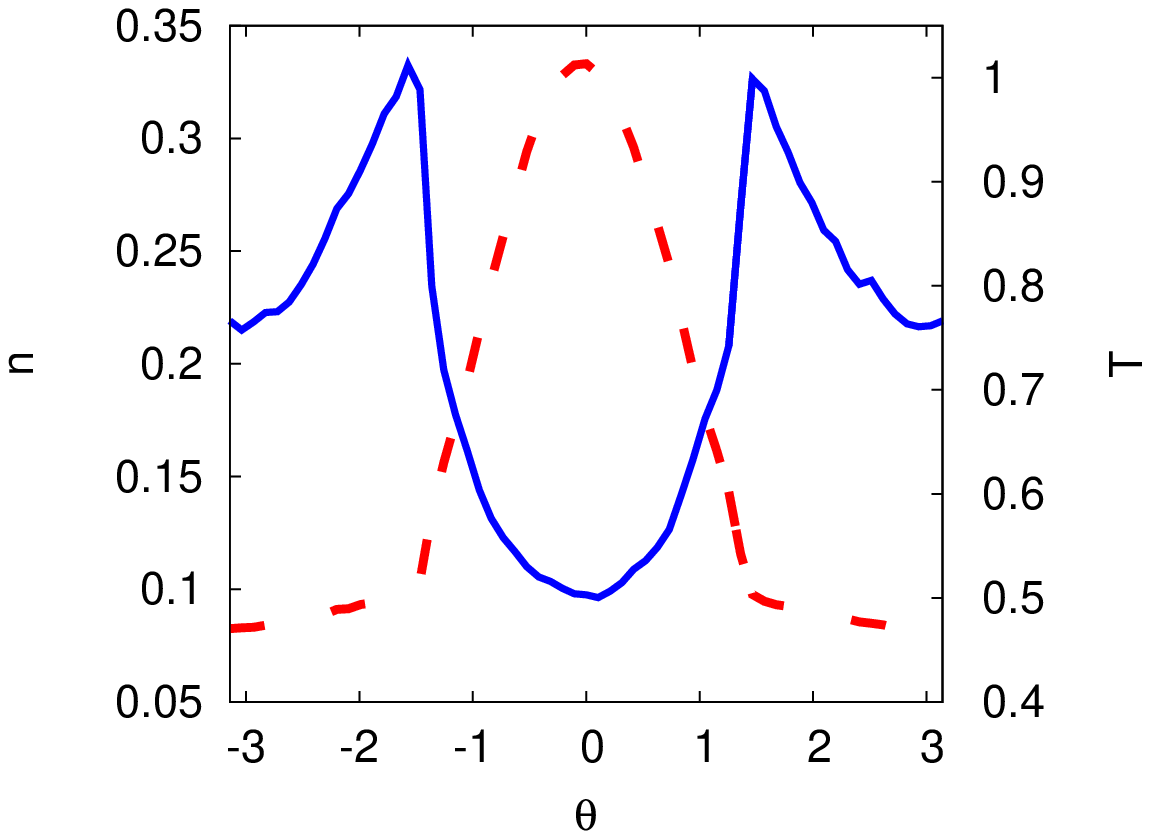}}
\caption{HMF: Temperature profile $T(\vartheta)$ (blue solid
line) and density profile $n(\vartheta)$ (red dashed line) measured in the
QSS at $t = 10^4$, in the same
conditions as Fig.\ 1 of the article, with the difference that $h=1.0$.}
\label{fig-sm:h1}
\end{figure}

\begin{figure}
\centerline{\includegraphics[width=65mm]{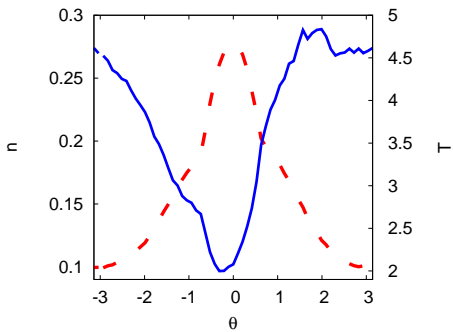}}
\caption{HMF: Temperature profile $T(\vartheta)$ (blue solid
line) and density profile $n(\vartheta)$ (red dashed line) measured in the
QSS at $t = 10^4$, in the same
conditions as Fig.\ 1 of the article, with the difference that the
duration of the impulse is $\tau=4$.}
\label{fig-sm:tau4}
\end{figure}

In Fig.\ 2 of the article, we have shown the density profile
$n(\vartheta)$ and temperature profile $T(\vartheta)$ in the QSS at $t =
10^4$, obtained with $m_0 = 0.521$ (corresponding to $T = 0.4244$), $t_0
= 100$, $h = 10$, and $\tau = 1$. In the following, and in particular,
in Figs. \ref{fig-sm:temp0-25}, \ref{fig-sm:h1}, and \ref{fig-sm:tau4},
we plot $n$ and $T$, still measured at $t = 10^4$ after a kick given at $t_0 = 100$, but with different values for other parameters, as described in the captions.

In Figs. \ref{fig-sm:temp0-25} and \ref{fig-sm:tau4}, we see temperature
inversion on the whole range of $\vartheta$'s. In Fig.\ \ref{fig-sm:h1},
inversion is present only on about half the interval, corresponding to
$-1.5 \lesssim \vartheta \lesssim 1.5$, where the state is sufficiently
clustered. In the rest of the interval, the density is almost flat, and there is no temperature inversion, although the temperature profile is still nonuniform. 

\begin{figure}
\centerline{\includegraphics[width=65mm]{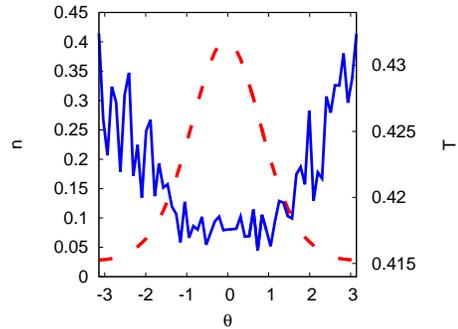}}
\caption{HMF: Temperature profile $T(\vartheta)$ (blue solid
line) and density profile $n(\vartheta)$ (red dashed line) measured in the
QSS at $t = 10^4$, while starting from a a Maxwellian velocity
distribution at temperature $T=0.4244$, but with a spatial distribution whose
magnetization is smaller than the corresponding equilibrium one by an amount $\Delta
m=0.05$.}
\label{fig-sm:dm}
\end{figure}

In Fig.\ \ref{fig-sm:dm}, we plot $n$ and $T$ at $t = 10^4$ as obtained
using a different protocol. Instead of preparing the system at thermal
equilibrium, and then adding an impulsive perturbation, we prepare the
system with a thermal velocity distribution at $T = 0.4244$, but with a
spatial distribution corresponding to a value $m$ of the magnetization
that it slightly smaller than the equilibrium value $m_0 = 0.521$, that
is, $m = 0.516$. Again, the system ends up in a QSS with temperature
inversion, but now the effect is less pronounced, since the state is much closer to an equilibrium one than in the other cases.

\subsubsection{Momentum distributions}

\begin{figure}
\centerline{\includegraphics[width=65mm]{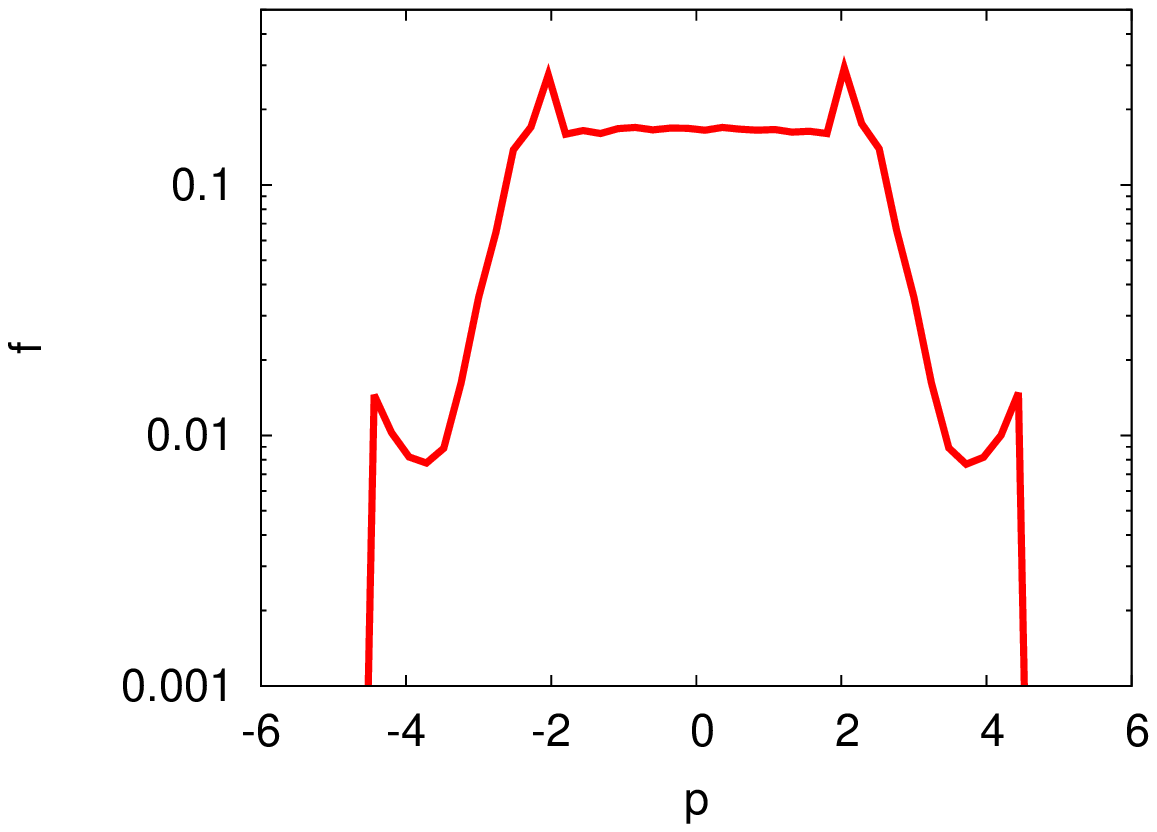}}
\centerline{\includegraphics[width=65mm]{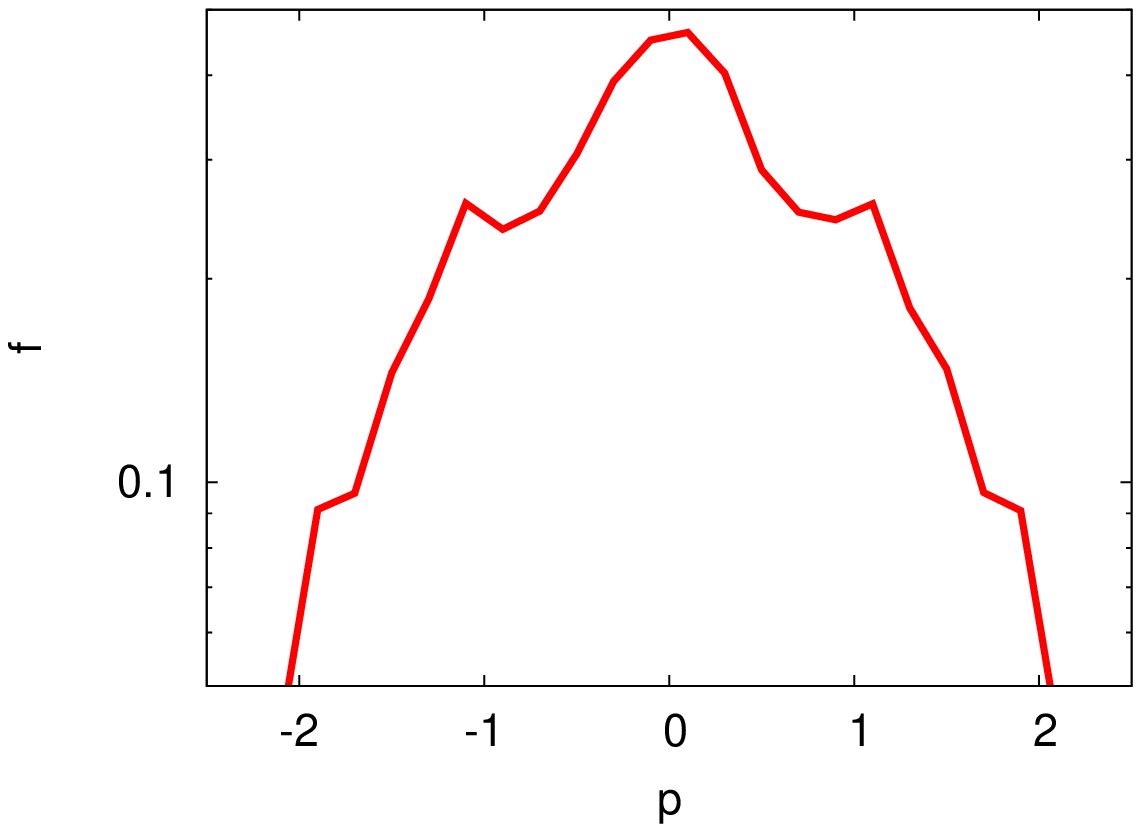}}
\caption{HMF: Momentum distribution at $t=10^4$, in the same
conditions as Fig.\ 1 of the main text, with the difference that the
impulsive perturbation is of the form \eqref{eq:vark} with $k = 2$ (top panel) and $k=8$ (bottom panel).}
\label{fig-sm:perturbation}
\end{figure}

In the article, it is argued that temperature inversion is produced by
velocity filtration after the perturbation has changed the initially
Maxwellian momentum distribution into a non-thermal one, due to Landau
damping. The fact that the energy exchange between the perturbation and
the particles occurs via a kind of wave-particle interaction is witnessed
by the fact that energy is preferentially absorbed by particles having
velocities close to some given values (see Fig.\ 4 of the article). As a
further support to this explanation, in Fig.\ \ref{fig-sm:perturbation}, we show that if we change the shape of the perturbation from that given in \eqref{eq:pert} into
\beq
\mathcal{H}_h \equiv -h\sum_{i=1}^N \cos (k\vartheta_i)
\label{eq:vark}
\eeq
with $k \not = 1$, we still have preferential absorption of energy for
specific values of the momentum, but these values are different from the case $k = 1$ and depend on $k$.

\subsection{Extended HMF model}

In order to show that including further Fourier components in the interaction potential does not destroy the phenomenon of temperature inversion, we now report on some results obtained for an extended HMF model, where also the second mode in the Fourier expansion of the potential is retained, defined by the Hamiltonian 
\beq
\label{hmf2}
\begin{split}
\mathcal{H}_{\text{HMF2}} = & \sum_{i=1}^{N} \frac{p_i^2}{2} + \frac{1}{N}\sum_{i = 1}^N \sum_{j > i}^N  \left[ 1 - K \cos \left(\vartheta_i - \vartheta_j \right) \right. \\
- & \left. \left(1 - K\right)\cos \left(2 \vartheta_i - 2 \vartheta_j \right) \right]\,.
\end{split}
\eeq
Note that for $K = 1$ one recovers the HMF model. 

\subsubsection{Density and temperature profiles}

\begin{figure}
\centerline{\includegraphics[width=65mm]{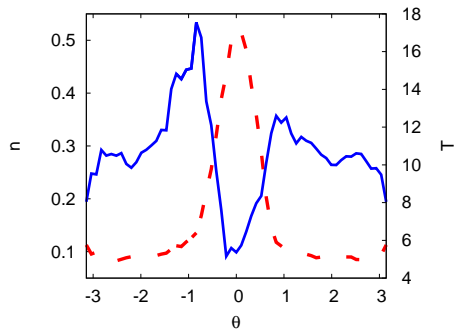}}
\caption{Extended HMF model: temperature profile $T(\vartheta)$ (blue solid
line) and density profile $n(\vartheta)$ (red dashed line) measured in the
QSS at $t = 10^3$, while starting from a thermal equilibrium state at temperature $T=0.25$ kicked out of equilibrium by an external field $h = 10$ acting for one time unit. Here $N = 10^7$.}
\label{fig-sm:2ndfourier}
\end{figure}

We considered the extended HMF model \eqref{hmf2} with $K = 0.3$. Also in this case, starting from equilibrium and kicking the system by switching on an external field for a short time we observed temperature inversion. In this case the inversion is limited to the region where the system is more collapsed, but still the temperature difference is large (see Fig.\ \ref{fig-sm:2ndfourier}).

\subsection{Two-dimensional self-gravitating system}

In the article we have also considered a two-dimensional self-gravitating system (2DSGS), that is, a system of $N$ point particles of mass $m$ moving in a plane with Hamiltonian 
\beq
\begin{split}
\mathcal{H}_{\text{G}} = & \sum_{i = 1}^N \frac{\left|\mathbf{p}_i\right|^2}{2m} \\
+ & \,\,  Gm^2
\sum_{i = 1}^N \sum_{j > i}^N 
\ln\left[\frac{\sqrt{(\mathbf{r}_i - \mathbf{r}_j)^2 + r_0^2}}{s}\right]\,,
\end{split}
\label{ham_2dgravApp}
\eeq
where $\mathbf{r}_i = (x_i,y_i)$, $s$ is a length scale, $\mathbf{p}_i = (\dot{x}_i,\dot{y}_i)$, $r_0$ is a small-scale cutoff to avoid the divergence when $\left| \mathbf{r}_i - \mathbf{r}_j \right| \to 0$, 
and $G$ is the gravitational constant. Such a system can be seen as a simple model of filamentary interstellar clouds. 

\subsubsection{Thermal equilibrium state}

The simulations reported in the article have been performed considering $N = 3 \times 10^4$ particles initially in a thermal equilibrium state, whose radial surface mass density profile is \cite{Ostriker:apj1964}
\beq
\varrho(r) = \frac{\varrho_0}{\left(1 + \frac{1}{8} x^2\right)^2}\,
\label{sgseqdensity}
\eeq 
where $x = r/s$ and $\varrho_0$ is the central surface density. By choosing
\beq
s = \sqrt{\frac{1}{4\pi G \beta \varrho_0}}
\label{s-def}
\eeq 
where 
\beq
\beta = \frac{m}{k_B T}
\eeq
is the parameter fixing the width of the equilibrium Maxwell distribution of the velocities and $k_B$ is the Boltzmann constant, one gets that $\beta$ depends only on the total mass (per unit length, since we are considering a two-dimensional slice of an infinite cylinder) $M$ of the system \cite{Ostriker:apj1964}:
\beq
\beta = \frac{2}{GM}\,.
\eeq
Note that $M$ is a finite quantity even if the system is unbounded in the plane due to the confining nature of the logarithmic two-dimensional gravitational potential. We used units where $G = M = 1$, and we also fixed $s = 1$, so that the equilibrium value of $\beta$ corresponding to the profile \eqref{sgseqdensity} is $\beta = 2$ and the central density is fixed by Eq.\ \eqref{s-def}, yielding $\varrho_0 = \left(4 \pi \right)^{-1}$ and a half-mass radius $x_* = r_*/s = \sqrt{8}$.

\subsubsection{Distribution functions}

\begin{figure}
\centerline{\includegraphics[width=65mm]{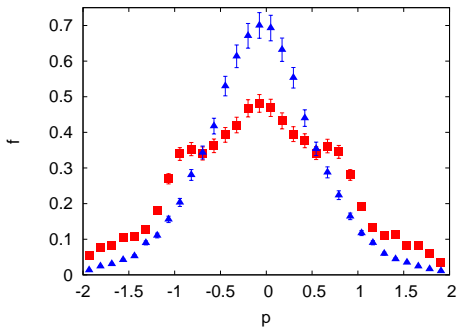}}
\caption{2DSGS: Distribution function $f(r,p_r)$ measured in the QSS as in Fig.\ 3 of the article, at $r = 7.9$ (red squares) and $r = 1.4$ (blue triangles).}
\label{sm:fpgravity}
\end{figure}

In the article we reported on simulations where, starting from the equilibrium described above, particles were kicked out of equilibrium by instantaneously adding to all their radial velocities the same amount $\delta v_r = \sigma_r/2$, where $\sigma_r$ is the radial velocity dispersion, that is, $\sigma_r = \beta^{-1/2}$. In those conditions we observed temperature inversion in the QSS up to radii larger than $r_*$. In Fig.\ \ref{sm:fpgravity} we plot the distribution function $f(r,p_r)$ measured in the QSS at two radii, that is, $r = 1.4$ where the temperature is close to its minimum and the density is close to its maximum, and $r = 7.9$ where the temperature is close to its maximum. The growth of the suprathermal tails in the less dense parts of the system as well as ``shoulders'' around $\left| p_r \right| \simeq 1$ are well apparent.  
 
\subsubsection{Other simulation protocols}

We also observed temperature inversion with other simulation protocols.

We found that temperature inversion also arises in long-lasting transients of the dissipationless collapse of initially ``cold'' (i.e., initial virial ratio $2K/|V| \leq 1$, where $K$ and $V$ are the totale kinetic and potential energies, repectively) two-dimensional self-gravitating systems. In this case, the initial condition for the numerical simulations was, as before, the density profile given in Eq.\ \eqref{sgseqdensity}, while velocities were extracted from a thermal distribution with $\beta=10$. Radial temperature profiles showing inversion up to the radius enclosing $70\%$ of the total mass $M$ last for times $t \simeq 20t_{dyn}$, after the violent relaxation phase, before the system settles to an isothermal profile with an abrupt fall at large radii. Perfectly cold initial conditions (i.e., $2K/|V|=0$ at $t=0$) were excluded from this study as they are known to be prone to the so-called \textit{bar instability} breaking the system's axial symmetry (see e.g.\ \cite{BinneyTremaine:book}).


\subsection{Velocity filtration and temperature inversion}

Velocity filtration was first suggested by Scudder
\cite{Scudder:apj1992a,Scudder:apj1992b,Scudder:apj1994} to explain the
heating of the solar corona. We now describe the original argument by
Scudder in the simple case of noninteracting particles in an external
potential, and then discuss why such a mechanism works also for a
long-range interacting system, as shown in \cite{epjb2014}. 

Let us consider a system of noninteracting particles moving in one spatial 
dimension on a semi-infinite line $x \geq 0$. The particles are subjected to an
external potential $\psi(x)$, such that the potential energy is growing
with $x$, $\psi'(x) > 0$. A typical case is that of an atmosphere, that
is, a gas of particles in a gravity field generated by a planet. In this
case, $x$ is the height above the ground, and the potential energy $\psi(x)$ is the gravitational one,
\beq
\psi(x) = \frac{GM}{R} \frac{x}{x+R}\, ,
\label{eq:gravpotential}
\eeq
where $G$ is the gravitational constant, $M$ is the mass of the planet, and $R$ its radius.
The one-particle distribution function $f(x,p,t)$ is a solution of the Vlasov equation
\beq
\frac{\partial f}{\partial t} + p \frac{\partial f}{\partial x} -
\frac{d\psi}{dx}\frac{\partial f}{\partial p} = 0\, ,
\label{vlasovApp}
\eeq 
with a given stationary boundary condition at $x = 0$, i.e.,
\beq
f_0(p) \equiv f(0,p,t)\, .
\eeq
The Vlasov equation (\ref{vlasovApp}) is nothing but energy conservation for
each particle, so that due to the potential $\psi(x)$, there
will be a ``velocity filtration'' effect. This means only those particles whose kinetic energy $k$ at $x = 0$ is sufficiently large to overcome the potential barrier $\Delta\psi(x') = \psi(x') - \psi(0)$  will reach the position $x'$ where their kinetic energy will be $k' = k - \Delta\psi(x')$. As a consequence, the spatial density
\beq
n(x) = \int_{-\infty}^{\infty} dp\, f(x,p)
\eeq
is a decreasing function of $x$. We now consider a Maxwellian boundary condition,
\beq
f_0^{\rm M}(p) = \frac{n_0}{\left(2\pi T_0 \right)^{1/2}} \exp\left(-\frac{p^2}{2T_0} \right)\, ,
\label{eq:maxwell}
\eeq
where $n_0 = n(0)$, and for simplicity we have taken particles of unit
mass and set the Boltzmann constant $k_B$ to unity. Then, the
stationary solution $f(x,p)$ of Eq.\ (\ref{vlasovApp}) is the so-called ``exponential
atmosphere'':
\beq
f(x,p) = \exp \left[- \frac{\psi(x) - \psi(0)}{T_0} \right] f_0^{\rm M}(p)\, .
\label{ea}
\eeq
The system is isothermal, i.e., the temperature profile
\beq
T(x) =  \frac{1}{n(x)}\int_{-\infty}^\infty dp\, p^2 f(x,p)  
\eeq
is constant, $T(x) = T_0$. In this case, the only effect of velocity filtration is that the density decreases with the height $x$. However, this is a very special case. Indeed, while velocity filtration occurs for any boundary condition $f_0(p)$, it yields a constant temperature profile only if the boundary condition is Maxwellian. This is due to the fact that $f(x,p)$ given by Eq.\ (\ref{ea}) is separable in $x$ and $p$. 

\begin{figure}
\centerline{\includegraphics[width=65mm]{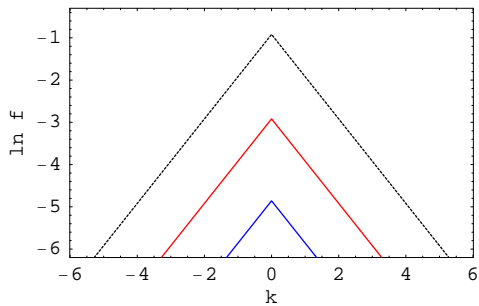}}
\caption{Logarithm of the distribution function $f(x,p)$ as a function of the signed kinetic energy $k$ calculated for noninteracting particles in a gravitational potential \eqref{eq:gravpotential} with $GM = 10$ and $R =1$ at height $x = 0$ (black dotted line), $x = 0.25$ (red line), and $x = 0.65$ (blue line). The black dotted line coincides with the boundary condition $f^0_M$ as given by Eq.\ (\ref{eq:maxwell}).}
\label{fig_scudder_maxw}
\end{figure}

\begin{figure}
\centerline{\includegraphics[width=65mm]{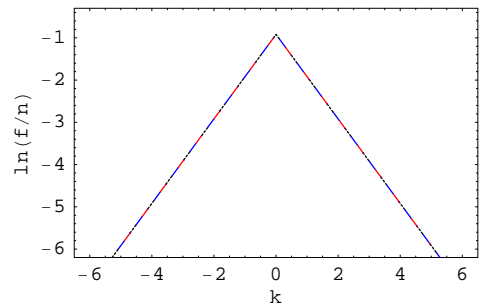}}
\caption{Logarithm of the distribution function divided by the local density, $f(x,p)/n(x)$, as a function of the signed kinetic energy $k$ in the same conditions as Fig.\ \ref{fig_scudder_maxw}.}
\label{fig_scudder_maxw_rescaled}
\end{figure}

In Ref.\ \cite{Scudder:apj1992a}, this was explained using a graphical construction. 
If we plot $\ln f(x,p)$ as a
function of the kinetic energy $k = p^2/2$, we get a straight line. Now
$f(x,k)$ is obtained from $f(0,k)$ by removing the part of the
distribution corresponding to kinetic energies smaller than the
potential barrier $\Delta\psi(x)$, and then rigidly translating the remaining
part towards the origin. We thus get a straight line with the same
slope, and since in this case $T = - (d\ln f/dk)^{-1}$, the temperature is the same at any $x$. In Fig.\ \ref{fig_scudder_maxw} we plot $\ln f$ as a function of the signed kinetic energy
\beq 
k = \text{sgn}(p) \frac{p^2}{2}
\eeq 
at different heights $x$ in a gravity field given by Eq.\ \eqref{eq:gravpotential} with $GM = 10$ and $R=1$, calculated solving Eq.\ (\ref{vlasovApp}) with the Maxwellian boundary condition $f_0^{\rm M}$ given in Eq.\ \eqref{eq:maxwell}. In Fig.\ \ref{fig_scudder_maxw_rescaled} we show that the curves collapse onto each other if rescaled by the density $n(x)$, as follows from Eq.\ \eqref{ea}.
      
\begin{figure}
\centerline{\includegraphics[width=65mm]{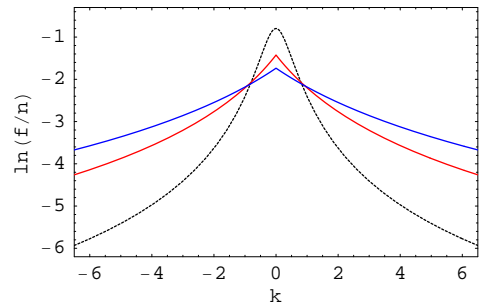}}
\caption{Logarithm of the distribution function normalized by the density,
$f(x,p)/n(x)$, calculated for noninteracting particles in the same gravity field as in Fig.\ \ref{fig_scudder_maxw}, at height $x = 0$ (black dotted line), $x = 0.25$ (red line), and $x = 0.65$ (blue line). The black dotted line coincides with the boundary condition $f_0(p) \propto p^{-4}$ as given by Eq.\ (\ref{f04}).}
\label{fig_scudder_f}
\end{figure}

\begin{figure}
\centerline{\includegraphics[width=65mm]{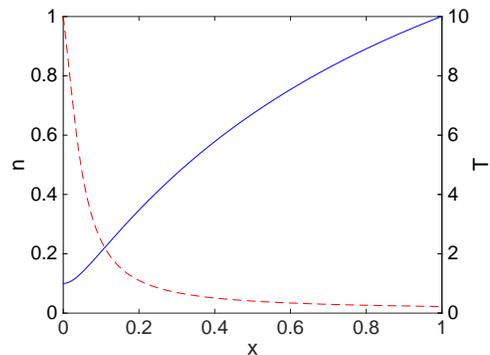}}
\caption{Temperature (blue solid line) and density (red dashed line) profiles calculated for noninteracting particles in a gravity field in the same conditions as Fig.\ \ref{fig_scudder_f}.}
\label{fig_scudder_calculation}
\end{figure}

However, the same reasoning implies that if the tails of $f_0(p)$ are more populated than in a Maxwellian (i.e., $f_0(p)$ is suprathermal),
then velocity filtration yields a broader distribution at $x$ than at the boundary. 
To give an example, in Fig.\ \ref{fig_scudder_f}, we plot the logarithm of distribution function divided by the density, $\ln \left[f(x,p)/n(x)\right]$, as a function of $k$ at different heights $x$ in the same gravity field as in Fig.\ \ref{fig_scudder_maxw}, with a power-law boundary condition
\beq
f_0(p) = \frac{\sqrt{2}}{\pi\left(1 + p^4 \right)}\,.
\label{f04}
\eeq
The broadening of the distribution as $x$ increases is well apparent, at variance with the collapse of the curves in the Maxwellian case, implying that the temperature is larger at larger heights. Indeed, $T$ grows as $n$ decreases, as shown in Fig.\ \ref{fig_scudder_calculation} where we plot the density and temperature profiles calculated in the same conditions as in Fig.\ \ref{fig_scudder_f}.

Why do we expect the same phenomenon to also occur in a
long-range-interacting system in an inhomogeneous QSS? Because for times
for which the system stays in a QSS, its one-particle distribution
function $f(\vartheta,p,t)$ still obeys a Vlasov equation,
\beq
\frac{\partial f}{\partial t} + p \frac{\partial f}{\partial \vartheta}
- \frac{\partial\left(\langle u \rangle + \psi\right)}{\partial \vartheta}\frac{\partial
f}{\partial p} = 0\, , 
\label{vlasov_lr}
\eeq 
where we have written $\vartheta$ instead of $x$ because we have in mind systems with periodic coordinates like the HMF model \eqref{hmfApp}.
The only difference with respect to Eq.\ (\ref{vlasovApp}) is that the
potential is now composed of two terms: the mean-field interaction,
\beq
\langle u \rangle (\vartheta,t) = \int d\vartheta' \int dp'\, u(\vartheta
- \vartheta') f(\vartheta',p',t), 
\label{u} 
\eeq
where $u$ is the two-body interaction potential, and (possibly) an external field $\psi(\vartheta)$, such that the total potential energy $V$ is 
\beq
V(\vartheta_1,\ldots,\vartheta_N) = \frac{1}{N}\sum_{i=1}^N \sum_{j < i}^N u(\vartheta_i - \vartheta_j)\, + \sum_{i=1}^N \psi(\vartheta_i)\,. 
\eeq
For the HMF model (\ref{hmfApp}), we have
\beq
u(\vartheta - \vartheta') =  1 - \cos(\vartheta - \vartheta')\, . 
\eeq
When present, the external potential term typically describes the interaction with a magnetic field $h$, so that it is given by
\beq
\psi(\vartheta) =  - h \cos \vartheta\,.
\eeq
Hence, despite the fact that each particle is
strongly interacting with all the other particles, the dynamics of our
system in a QSS is equivalent to that of noninteracting particles
immersed in a self-consistent field given by $\langle u \rangle + \psi$.
From the above discussion, it should be clear that the key ingredients
to obtain a temperature inversion due to velocity filtration in the
noninteracting particle case are $(i)$ the presence of an attractive
external field, and $(ii)$ the fact that $f(p)$ is suprathermal. Hence,
we should expect the same behavior also in the long-range-interacting
case in a QSS, provided the resulting effective field $\langle u
\rangle + \psi$ is attractive---or, equivalently, the density profile is
clustered due to attractive self-interactions---and $f(p)$ is suprathermal.

\bibliography{temp_inv_lr_transfer}

\begin{thebibliography}{37}%
\makeatletter
\providecommand \@ifxundefined [1]{%
 \@ifx{#1\undefined}
}%
\providecommand \@ifnum [1]{%
 \ifnum #1\expandafter \@firstoftwo
 \else \expandafter \@secondoftwo
 \fi
}%
\providecommand \@ifx [1]{%
 \ifx #1\expandafter \@firstoftwo
 \else \expandafter \@secondoftwo
 \fi
}%
\providecommand \natexlab [1]{#1}%
\providecommand \enquote  [1]{``#1''}%
\providecommand \bibnamefont  [1]{#1}%
\providecommand \bibfnamefont [1]{#1}%
\providecommand \citenamefont [1]{#1}%
\providecommand \href@noop [0]{\@secondoftwo}%
\providecommand \href [0]{\begingroup \@sanitize@url \@href}%
\providecommand \@href[1]{\@@startlink{#1}\@@href}%
\providecommand \@@href[1]{\endgroup#1\@@endlink}%
\providecommand \@sanitize@url [0]{\catcode `\\12\catcode `\$12\catcode
  `\&12\catcode `\#12\catcode `\^12\catcode `\_12\catcode `\%12\relax}%
\providecommand \@@startlink[1]{}%
\providecommand \@@endlink[0]{}%
\providecommand \url  [0]{\begingroup\@sanitize@url \@url }%
\providecommand \@url [1]{\endgroup\@href {#1}{\urlprefix }}%
\providecommand \urlprefix  [0]{URL }%
\providecommand \Eprint [0]{\href }%
\providecommand \doibase [0]{http://dx.doi.org/}%
\providecommand \selectlanguage [0]{\@gobble}%
\providecommand \bibinfo  [0]{\@secondoftwo}%
\providecommand \bibfield  [0]{\@secondoftwo}%
\providecommand \translation [1]{[#1]}%
\providecommand \BibitemOpen [0]{}%
\providecommand \bibitemStop [0]{}%
\providecommand \bibitemNoStop [0]{.\EOS\space}%
\providecommand \EOS [0]{\spacefactor3000\relax}%
\providecommand \BibitemShut  [1]{\csname bibitem#1\endcsname}%
\let\auto@bib@innerbib\@empty
\bibitem [{Note1()}]{Note1}%
  \BibitemOpen
  \bibinfo {note} {Interactions are called long-range when they decay
  asymptotically with the interparticle distance $r$ as $r^{-\alpha }$ with $0
  \leq \alpha \leq d$ in $d$ dimensions \cite {CampaEtAl:book}.}\BibitemShut
  {Stop}%
\bibitem [{\citenamefont {Campa}\ \emph {et~al.}(2014)\citenamefont {Campa},
  \citenamefont {Dauxois}, \citenamefont {Fanelli},\ and\ \citenamefont
  {Ruffo}}]{CampaEtAl:book}%
  \BibitemOpen
  \bibfield  {author} {\bibinfo {author} {\bibfnamefont {A.}~\bibnamefont
  {Campa}}, \bibinfo {author} {\bibfnamefont {T.}~\bibnamefont {Dauxois}},
  \bibinfo {author} {\bibfnamefont {D.}~\bibnamefont {Fanelli}}, \ and\
  \bibinfo {author} {\bibfnamefont {S.}~\bibnamefont {Ruffo}},\ }\href@noop {}
  {\emph {\bibinfo {title} {Physics of Long-Range Interacting Systems}}}\
  (\bibinfo  {publisher} {Oxford University Press},\ \bibinfo {address}
  {Oxford},\ \bibinfo {year} {2014})\BibitemShut {NoStop}%
\bibitem [{\citenamefont {Choudouri}(2010)}]{Choudhouri:book}%
  \BibitemOpen
  \bibfield  {author} {\bibinfo {author} {\bibfnamefont {A.~R.}\ \bibnamefont
  {Choudouri}},\ }\href@noop {} {\emph {\bibinfo {title} {Astrophysics for
  Physicists}}}\ (\bibinfo  {publisher} {Cambridge University Press},\ \bibinfo
  {address} {Cambridge},\ \bibinfo {year} {2010})\BibitemShut {NoStop}%
\bibitem [{\citenamefont {Binney}\ and\ \citenamefont
  {Tremaine}(2008)}]{BinneyTremaine:book}%
  \BibitemOpen
  \bibfield  {author} {\bibinfo {author} {\bibfnamefont {J.}~\bibnamefont
  {Binney}}\ and\ \bibinfo {author} {\bibfnamefont {S.}~\bibnamefont
  {Tremaine}},\ }\href@noop {} {\emph {\bibinfo {title} {Galactic Dynamics}}},\
  \bibinfo {edition} {2nd}\ ed.\ (\bibinfo  {publisher} {Princeton University
  Press},\ \bibinfo {address} {Princeton},\ \bibinfo {year} {2008})\BibitemShut
  {NoStop}%
\bibitem [{Note2()}]{Note2}%
  \BibitemOpen
  \bibinfo {note} {Also the related issue of \protect \textit {stability} of
  the Vlasov stationary states is resolved only in particular cases \cite
  {CampaEtAl:book}.}\BibitemShut {Stop}%
\bibitem [{\citenamefont {Lynden-Bell}(1967)}]{Lynden-Bell:mnras1967}%
  \BibitemOpen
  \bibfield  {author} {\bibinfo {author} {\bibfnamefont {D.}~\bibnamefont
  {Lynden-Bell}},\ }\href@noop {} {\bibfield  {journal} {\bibinfo  {journal}
  {Monthly Notices of the Royal Astronomical Society}\ }\textbf {\bibinfo
  {volume} {136}},\ \bibinfo {pages} {101} (\bibinfo {year}
  {1967})}\BibitemShut {NoStop}%
\bibitem [{\citenamefont {Levin}\ \emph {et~al.}(2014)\citenamefont {Levin},
  \citenamefont {Pakter}, \citenamefont {Rizzato}, \citenamefont {Teles},\ and\
  \citenamefont {Benetti}}]{LevinEtAlphysrep:2014}%
  \BibitemOpen
  \bibfield  {author} {\bibinfo {author} {\bibfnamefont {Y.}~\bibnamefont
  {Levin}}, \bibinfo {author} {\bibfnamefont {R.}~\bibnamefont {Pakter}},
  \bibinfo {author} {\bibfnamefont {F.~B.}\ \bibnamefont {Rizzato}}, \bibinfo
  {author} {\bibfnamefont {T.~N.}\ \bibnamefont {Teles}}, \ and\ \bibinfo
  {author} {\bibfnamefont {F.~P.~C.}\ \bibnamefont {Benetti}},\ }\href
  {\doibase http://dx.doi.org/10.1016/j.physrep.2013.10.001} {\bibfield
  {journal} {\bibinfo  {journal} {Physics Reports}\ }\textbf {\bibinfo {volume}
  {535}},\ \bibinfo {pages} {1 } (\bibinfo {year} {2014})}\BibitemShut
  {NoStop}%
\bibitem [{\citenamefont {Benetti}\ \emph {et~al.}(2014)\citenamefont
  {Benetti}, \citenamefont {Ribeiro-Teixeira}, \citenamefont {Pakter},\ and\
  \citenamefont {Levin}}]{BenettiEtAl:prl2014}%
  \BibitemOpen
  \bibfield  {author} {\bibinfo {author} {\bibfnamefont {F.~P.~C.}\
  \bibnamefont {Benetti}}, \bibinfo {author} {\bibfnamefont {A.~C.}\
  \bibnamefont {Ribeiro-Teixeira}}, \bibinfo {author} {\bibfnamefont
  {R.}~\bibnamefont {Pakter}}, \ and\ \bibinfo {author} {\bibfnamefont
  {Y.}~\bibnamefont {Levin}},\ }\href {\doibase 10.1103/PhysRevLett.113.100602}
  {\bibfield  {journal} {\bibinfo  {journal} {Phys. Rev. Lett.}\ }\textbf
  {\bibinfo {volume} {113}},\ \bibinfo {pages} {100602} (\bibinfo {year}
  {2014})}\BibitemShut {NoStop}%
\bibitem [{Note3()}]{Note3}%
  \BibitemOpen
  \bibinfo {note} {At variance with short-range systems, equilibrium states of
  long-range systems can be inhomogeneous, typically when the interactions are
  attractive.}\BibitemShut {Stop}%
\bibitem [{Note4()}]{Note4}%
  \BibitemOpen
  \bibinfo {note} {Temperature inversions have been observed also in other
  systems like the hot gas of ``cooling-core'' galaxy clusters \cite
  {WiseMcNamaraMurray:apj2004} and the Io plasma torus around Jupiter \cite
  {MeyerVernetMoncuquetHoang:icarus1995}.}\BibitemShut {Stop}%
\bibitem [{\citenamefont {Myers}\ and\ \citenamefont
  {Fuller}(1992)}]{MyersFuller:apj1992}%
  \BibitemOpen
  \bibfield  {author} {\bibinfo {author} {\bibfnamefont {P.~C.}\ \bibnamefont
  {Myers}}\ and\ \bibinfo {author} {\bibfnamefont {G.~A.}\ \bibnamefont
  {Fuller}},\ }\href@noop {} {\bibfield  {journal} {\bibinfo  {journal} {The
  Astrophysical Journal}\ }\textbf {\bibinfo {volume} {396}},\ \bibinfo {pages}
  {631} (\bibinfo {year} {1992})}\BibitemShut {NoStop}%
\bibitem [{\citenamefont {Goodman}\ \emph {et~al.}(1998)\citenamefont {Goodman}
  \emph {et~al.}}]{GoodmanEtAl:apj1998}%
  \BibitemOpen
  \bibfield  {author} {\bibinfo {author} {\bibfnamefont {A.~A.}\ \bibnamefont
  {Goodman}} \emph {et~al.},\ }\href {\doibase 10.1086/306045} {\bibfield
  {journal} {\bibinfo  {journal} {The Astrophysical Journal}\ }\textbf
  {\bibinfo {volume} {504}},\ \bibinfo {pages} {223} (\bibinfo {year}
  {1998})}\BibitemShut {NoStop}%
\bibitem [{\citenamefont {Pineda}\ \emph {et~al.}(2010)\citenamefont {Pineda}
  \emph {et~al.}}]{PinedaEtAl:apjl2010}%
  \BibitemOpen
  \bibfield  {author} {\bibinfo {author} {\bibfnamefont {J.~E.}\ \bibnamefont
  {Pineda}} \emph {et~al.},\ }\href {\doibase 10.1088/2041-8205/712/1/L116}
  {\bibfield  {journal} {\bibinfo  {journal} {The Astrophysical Journal
  Letters}\ }\textbf {\bibinfo {volume} {712}},\ \bibinfo {pages} {L116}
  (\bibinfo {year} {2010})}\BibitemShut {NoStop}%
\bibitem [{\citenamefont {Golub}\ and\ \citenamefont
  {Pasachoff}(2009)}]{GolubPasachoff:book}%
  \BibitemOpen
  \bibfield  {author} {\bibinfo {author} {\bibfnamefont {L.}~\bibnamefont
  {Golub}}\ and\ \bibinfo {author} {\bibfnamefont {J.~M.}\ \bibnamefont
  {Pasachoff}},\ }\href@noop {} {\emph {\bibinfo {title} {The Solar Corona}}},\
  \bibinfo {edition} {2nd}\ ed.\ (\bibinfo  {publisher} {Cambridge University
  Press},\ \bibinfo {address} {Cambridge},\ \bibinfo {year} {2009})\BibitemShut
  {NoStop}%
\bibitem [{\citenamefont {Aschwanden}(2005)}]{Aschwanden:book}%
  \BibitemOpen
  \bibfield  {author} {\bibinfo {author} {\bibfnamefont {M.~J.}\ \bibnamefont
  {Aschwanden}},\ }\href@noop {} {\emph {\bibinfo {title} {Physics of the Solar
  Corona. An Introduction with Problems and Solutions}}},\ \bibinfo {edition}
  {2nd}\ ed.\ (\bibinfo  {publisher} {Springer},\ \bibinfo {address} {New
  York},\ \bibinfo {year} {2005})\BibitemShut {NoStop}%
\bibitem [{\citenamefont {Klimchuk}(2006)}]{Klimchuk:sp2006}%
  \BibitemOpen
  \bibfield  {author} {\bibinfo {author} {\bibfnamefont {J.~A.}\ \bibnamefont
  {Klimchuk}},\ }\href {\doibase 10.1007/s11207-006-0055-z} {\bibfield
  {journal} {\bibinfo  {journal} {Solar Physics}\ }\textbf {\bibinfo {volume}
  {234}},\ \bibinfo {pages} {41} (\bibinfo {year} {2006})}\BibitemShut
  {NoStop}%
\bibitem [{Note5()}]{Note5}%
  \BibitemOpen
  \bibinfo {note} {By means of, e.g., magnetic reconnection or Alfv\'en waves
  \cite {Klimchuk:sp2006}.}\BibitemShut {Stop}%
\bibitem [{\citenamefont {Scudder}(1992{\natexlab{a}})}]{Scudder:apj1992a}%
  \BibitemOpen
  \bibfield  {author} {\bibinfo {author} {\bibfnamefont {J.~D.}\ \bibnamefont
  {Scudder}},\ }\href@noop {} {\bibfield  {journal} {\bibinfo  {journal} {The
  Astrophysical Journal}\ }\textbf {\bibinfo {volume} {398}},\ \bibinfo {pages}
  {299} (\bibinfo {year} {1992}{\natexlab{a}})}\BibitemShut {NoStop}%
\bibitem [{\citenamefont {Scudder}(1992{\natexlab{b}})}]{Scudder:apj1992b}%
  \BibitemOpen
  \bibfield  {author} {\bibinfo {author} {\bibfnamefont {J.~D.}\ \bibnamefont
  {Scudder}},\ }\href@noop {} {\bibfield  {journal} {\bibinfo  {journal} {The
  Astrophysical Journal}\ }\textbf {\bibinfo {volume} {398}},\ \bibinfo {pages}
  {319} (\bibinfo {year} {1992}{\natexlab{b}})}\BibitemShut {NoStop}%
\bibitem [{\citenamefont {Scudder}(1994)}]{Scudder:apj1994}%
  \BibitemOpen
  \bibfield  {author} {\bibinfo {author} {\bibfnamefont {J.~D.}\ \bibnamefont
  {Scudder}},\ }\href@noop {} {\bibfield  {journal} {\bibinfo  {journal} {The
  Astrophysical Journal}\ }\textbf {\bibinfo {volume} {427}},\ \bibinfo {pages}
  {446} (\bibinfo {year} {1994})}\BibitemShut {NoStop}%
\bibitem [{\citenamefont {Casetti}\ and\ \citenamefont
  {Gupta}(2014)}]{epjb2014}%
  \BibitemOpen
  \bibfield  {author} {\bibinfo {author} {\bibfnamefont {L.}~\bibnamefont
  {Casetti}}\ and\ \bibinfo {author} {\bibfnamefont {S.}~\bibnamefont
  {Gupta}},\ }\href {\doibase 10.1140/epjb/e2014-50136-y} {\bibfield  {journal}
  {\bibinfo  {journal} {The European Physical Journal B}\ }\textbf {\bibinfo
  {volume} {87}},\ \bibinfo {pages} {91} (\bibinfo {year} {2014})}\BibitemShut
  {NoStop}%
\bibitem [{\citenamefont {Anderson}(1994)}]{Anderson:apj1994}%
  \BibitemOpen
  \bibfield  {author} {\bibinfo {author} {\bibfnamefont {S.~W.}\ \bibnamefont
  {Anderson}},\ }\href@noop {} {\bibfield  {journal} {\bibinfo  {journal} {The
  Astrophysical Journal}\ }\textbf {\bibinfo {volume} {437}},\ \bibinfo {pages}
  {860} (\bibinfo {year} {1994})}\BibitemShut {NoStop}%
\bibitem [{\citenamefont {Landi}\ and\ \citenamefont
  {Pantellini}(2001)}]{LandiPantellini:aa2001}%
  \BibitemOpen
  \bibfield  {author} {\bibinfo {author} {\bibfnamefont {S.}~\bibnamefont
  {Landi}}\ and\ \bibinfo {author} {\bibfnamefont {F.~G.~E.}\ \bibnamefont
  {Pantellini}},\ }\href {\doibase 10.1051/0004-6361:20010552} {\bibfield
  {journal} {\bibinfo  {journal} {Astronomy \& Astrophysics}\ }\textbf
  {\bibinfo {volume} {372}},\ \bibinfo {pages} {686} (\bibinfo {year}
  {2001})}\BibitemShut {NoStop}%
\bibitem [{Note6()}]{Note6}%
  \BibitemOpen
  \bibinfo {note} {At variance with Scudder's model we do not need any special
  boundary condition.}\BibitemShut {Stop}%
\bibitem [{\citenamefont {Antoni}\ and\ \citenamefont
  {Ruffo}(1995)}]{AntoniRuffo:pre1995}%
  \BibitemOpen
  \bibfield  {author} {\bibinfo {author} {\bibfnamefont {M.}~\bibnamefont
  {Antoni}}\ and\ \bibinfo {author} {\bibfnamefont {S.}~\bibnamefont {Ruffo}},\
  }\href {\doibase 10.1103/PhysRevE.52.2361} {\bibfield  {journal} {\bibinfo
  {journal} {Phys. Rev. E}\ }\textbf {\bibinfo {volume} {52}},\ \bibinfo
  {pages} {2361} (\bibinfo {year} {1995})}\BibitemShut {NoStop}%
\bibitem [{Note7()}]{Note7}%
  \BibitemOpen
  \bibinfo {note} {The HMF may also be seen as a system of mean-field $XY$
  spins.}\BibitemShut {Stop}%
\bibitem [{Note8()}]{Note8}%
  \BibitemOpen
  \bibinfo {note} {We used a $4^{\protect \text {th}}$ order symplectic
  algorithm with time step $\delta t =0.1$, keeping relative energy
  fluctuations below $10^{-8}$.}\BibitemShut {Stop}%
\bibitem [{\citenamefont {de~Buyl}\ \emph {et~al.}(2011)\citenamefont
  {de~Buyl}, \citenamefont {Mukamel},\ and\ \citenamefont
  {Ruffo}}]{deBuylMukamelRuffo:pre2011}%
  \BibitemOpen
  \bibfield  {author} {\bibinfo {author} {\bibfnamefont {P.}~\bibnamefont
  {de~Buyl}}, \bibinfo {author} {\bibfnamefont {D.}~\bibnamefont {Mukamel}}, \
  and\ \bibinfo {author} {\bibfnamefont {S.}~\bibnamefont {Ruffo}},\ }\href
  {\doibase 10.1103/PhysRevE.84.061151} {\bibfield  {journal} {\bibinfo
  {journal} {Phys. Rev. E}\ }\textbf {\bibinfo {volume} {84}},\ \bibinfo
  {pages} {061151} (\bibinfo {year} {2011})}\BibitemShut {NoStop}%
\bibitem [{\citenamefont {Toci}\ and\ \citenamefont
  {Galli}(2015)}]{TociGalli:mnras2015}%
  \BibitemOpen
  \bibfield  {author} {\bibinfo {author} {\bibfnamefont {C.}~\bibnamefont
  {Toci}}\ and\ \bibinfo {author} {\bibfnamefont {D.}~\bibnamefont {Galli}},\
  }\href {\doibase 10.1093/mnras/stu2168} {\bibfield  {journal} {\bibinfo
  {journal} {Monthly Notices of the Royal Astronomical Society}\ }\textbf
  {\bibinfo {volume} {446}},\ \bibinfo {pages} {2110} (\bibinfo {year}
  {2015})}\BibitemShut {NoStop}%
\bibitem [{\citenamefont {Ostriker}(1964)}]{Ostriker:apj1964}%
  \BibitemOpen
  \bibfield  {author} {\bibinfo {author} {\bibfnamefont {J.}~\bibnamefont
  {Ostriker}},\ }\href@noop {} {\bibfield  {journal} {\bibinfo  {journal} {The
  Astrophysical Journal}\ }\textbf {\bibinfo {volume} {140}},\ \bibinfo {pages}
  {1056} (\bibinfo {year} {1964})}\BibitemShut {NoStop}%
\bibitem [{\citenamefont {Choudouri}(1998)}]{Choudhouri:plasmabook}%
  \BibitemOpen
  \bibfield  {author} {\bibinfo {author} {\bibfnamefont {A.~R.}\ \bibnamefont
  {Choudouri}},\ }\href@noop {} {\emph {\bibinfo {title} {The Physics of Fluids
  and Plasmas}}}\ (\bibinfo  {publisher} {Cambridge University Press},\
  \bibinfo {address} {Cambridge},\ \bibinfo {year} {1998})\BibitemShut
  {NoStop}%
\bibitem [{\citenamefont {Mohout}\ and\ \citenamefont
  {Villani}(2011)}]{MohoutVillani:actamath2011}%
  \BibitemOpen
  \bibfield  {author} {\bibinfo {author} {\bibfnamefont {C.}~\bibnamefont
  {Mohout}}\ and\ \bibinfo {author} {\bibfnamefont {C.}~\bibnamefont
  {Villani}},\ }\href {\doibase 10.1007/s11511-011-0068-9} {\bibfield
  {journal} {\bibinfo  {journal} {Acta Mathematica}\ }\textbf {\bibinfo
  {volume} {207}},\ \bibinfo {pages} {29} (\bibinfo {year} {2011})}\BibitemShut
  {NoStop}%
\bibitem [{\citenamefont {Barr\'{e}}\ \emph {et~al.}(2010)\citenamefont
  {Barr\'{e}}, \citenamefont {Olivetti},\ and\ \citenamefont
  {Yamaguchi}}]{BarreOlivettiYamaguchi:jstat2010}%
  \BibitemOpen
  \bibfield  {author} {\bibinfo {author} {\bibfnamefont {J.}~\bibnamefont
  {Barr\'{e}}}, \bibinfo {author} {\bibfnamefont {A.}~\bibnamefont {Olivetti}},
  \ and\ \bibinfo {author} {\bibfnamefont {Y.~Y.}\ \bibnamefont {Yamaguchi}},\
  }\href {\doibase 10.1088/1751-8113/44/40/405502} {\bibfield  {journal}
  {\bibinfo  {journal} {Journal of Statistical Mechanics: Theory and
  Experiment}\ }\textbf {\bibinfo {volume} {2010}},\ \bibinfo {pages} {P08002}
  (\bibinfo {year} {2010})}\BibitemShut {NoStop}%
\bibitem [{\citenamefont {Barr\'{e}}\ \emph {et~al.}(2011)\citenamefont
  {Barr\'{e}}, \citenamefont {Olivetti},\ and\ \citenamefont
  {Yamaguchi}}]{BarreOlivettiYamaguchi:jphysa2011}%
  \BibitemOpen
  \bibfield  {author} {\bibinfo {author} {\bibfnamefont {J.}~\bibnamefont
  {Barr\'{e}}}, \bibinfo {author} {\bibfnamefont {A.}~\bibnamefont {Olivetti}},
  \ and\ \bibinfo {author} {\bibfnamefont {Y.~Y.}\ \bibnamefont {Yamaguchi}},\
  }\href {\doibase 10.1088/1742-5468/2010/08/P08002} {\bibfield  {journal}
  {\bibinfo  {journal} {Journal of Physics A: Mathematical and Theoretical}\
  }\textbf {\bibinfo {volume} {44}},\ \bibinfo {pages} {405502} (\bibinfo
  {year} {2011})}\BibitemShut {NoStop}%
\bibitem [{\citenamefont {Bowers}\ and\ \citenamefont
  {Birdsall}(2002)}]{BowersBirdsall:physplasmas2002}%
  \BibitemOpen
  \bibfield  {author} {\bibinfo {author} {\bibfnamefont {K.~J.}\ \bibnamefont
  {Bowers}}\ and\ \bibinfo {author} {\bibfnamefont {C.~K.}\ \bibnamefont
  {Birdsall}},\ }\href@noop {} {\bibfield  {journal} {\bibinfo  {journal}
  {Physics of Plasmas}\ }\textbf {\bibinfo {volume} {9}},\ \bibinfo {pages}
  {2405} (\bibinfo {year} {2002})}\BibitemShut {NoStop}%
\bibitem [{\citenamefont {Wise}\ \emph {et~al.}(2004)\citenamefont {Wise},
  \citenamefont {McNamara},\ and\ \citenamefont
  {Murray}}]{WiseMcNamaraMurray:apj2004}%
  \BibitemOpen
  \bibfield  {author} {\bibinfo {author} {\bibfnamefont {M.~W.}\ \bibnamefont
  {Wise}}, \bibinfo {author} {\bibfnamefont {B.~R.}\ \bibnamefont {McNamara}},
  \ and\ \bibinfo {author} {\bibfnamefont {S.~S.}\ \bibnamefont {Murray}},\
  }\href@noop {} {\bibfield  {journal} {\bibinfo  {journal} {The Astrophysical
  Journal}\ }\textbf {\bibinfo {volume} {601}},\ \bibinfo {pages} {184}
  (\bibinfo {year} {2004})}\BibitemShut {NoStop}%
\bibitem [{\citenamefont {Meyer-Vernet}\ \emph {et~al.}(1995)\citenamefont
  {Meyer-Vernet}, \citenamefont {Moncuquet},\ and\ \citenamefont
  {Hoang}}]{MeyerVernetMoncuquetHoang:icarus1995}%
  \BibitemOpen
  \bibfield  {author} {\bibinfo {author} {\bibfnamefont {N.}~\bibnamefont
  {Meyer-Vernet}}, \bibinfo {author} {\bibfnamefont {M.}~\bibnamefont
  {Moncuquet}}, \ and\ \bibinfo {author} {\bibfnamefont {S.}~\bibnamefont
  {Hoang}},\ }\href {\doibase 10.1006/icar.1995.1121} {\bibfield  {journal}
  {\bibinfo  {journal} {Icarus}\ }\textbf {\bibinfo {volume} {116}},\ \bibinfo
  {pages} {202} (\bibinfo {year} {1995})}\BibitemShut {NoStop}%
\end{thebibliography}%

\end{document}